\documentclass[aps,prb,showpacs,reprint,superscriptaddress]{revtex4-2}

\usepackage{graphicx}
\usepackage{amsmath}
\usepackage{amssymb}
\usepackage{float}
\usepackage{bbm}
\usepackage{dsfont}
\usepackage{color}

\usepackage[utf8x]{inputenc}   % UTF-8 encoding
\usepackage[T1]{fontenc}      % T1 font encoding for special characters

\input epsf   % Only include if you're using EPS figures

\usepackage{verbatim}

\begin{document}

\title{Optimal superconductivity in twisted bilayer WSe$_2$ where the Van Hove singularity crosses half-filling}
\author{Michał Zegrodnik}
\email{michal.zegrodnik@agh.edu.pl}
\affiliation{Academic Centre for Materials and Nanotechnology, AGH University of Krakow, Al. Mickiewicza 30, 30-059 Krakow,
Poland}
\author{Waseem Akbar}
\affiliation{Academic Centre for Materials and Nanotechnology, AGH University of Krakow, Al. Mickiewicza 30, 30-059 Krakow,
Poland}
\author{Andrzej Biborski}
\affiliation{Academic Centre for Materials and Nanotechnology, AGH University of Krakow, Al. Mickiewicza 30, 30-059 Krakow,
Poland}
\author{Louk Rademaker}
\affiliation{Department of Theoretical Physics, University of Geneva, CH-1211 Geneva, Switzerland}
\affiliation{Institute-Lorentz for Theoretical Physics, Leiden University, PO Box 9506, Leiden, NL-2300, The Netherlands} 

%\date{06.05.2022}

\begin{abstract}
The recent discovery of unconventional superconductivity has pointed to twisted WSe$_2$ bilayer as a versatile platform for studying the correlated and topological phases of matter. Here we analyze the effect of the displacement field and electron interactions on the formation of a topological paired state in twisted WSe$_2$. Our approach is based on the effective single band $t$-$J$-$U$ model supplemented with intersite Coulomb interaction term and treated within the Gutzwiller approximation. 
We show that the superconducting phase is stabilized in a small range of displacement fields where the Van Hove singularity crosses half-filling, which is in qualitative agreement with recent experimental data. According to our analysis, such a circumstance comes as a result of a subtle interplay between the large density of states of the Van Hove singularity, in combination with the renormalization effects that appear in the weak-to-moderate correlations regime. The two factors create favorable conditions for the SC pairing only in a small area of the phase diagram.
 \end{abstract}
\maketitle

%%%%%%%%%%%%%%%%%%%%%%%%%%%%%%%%%%%%%%%%%%%%%%%%%%%%%%%%%%%%%%%%%%%%%%%%

%%%%%%%%%%%%%%%%%%%%%%%%%%%%%%%%%%%%%%%%%%%%%%%%%%%%%%%%%%%%%%%%%%%%%%%%
\textit{Introduction}.---Transition-metal dichalcogenide (TMD) bilayers have recently attracted a significant amount of interest due to their rich physics and exceptional tunability. It has been established experiementally that those systems host a number of correlated and topological states such as: Mott insulators~\cite{Ghiotto2021,Wang2020}, magnetically ordered states~\cite{Xu2023,Jie2024}, generalized Wigner crystals~\cite{Regan2020,Xu2020,Huang2021,Li2021,Zhou2024}, quantum anomalous and spin quantum anomalous Hall states~\cite{mw_exp_1,mw_exp_2,Tao2024}, as well as Kondo effects~\cite{Zhao2023}. Moreover, in recent years signatures of unconventional superconducting (SC) state have been reported, first by Wang et al. for the case of the twisted WSe$_2$ bilayer (tWSe$_2$)~\cite{Wang2020}. It has been suggested that for this system two superconducting domes reside on both sides of a Mott insulating state, which is located at half-filling similarly as in the well-known cuprates~\cite{Spalek2022} or twisted bilayer graphene~\cite{Cao2018_1,Cao2018_2}. Interestingly, strong evidence of the paired state in tWSe$_2$ has been shown in very recent experimental papers, which, however, report one SC dome close to or at the half-filling for a certain range of displacement fields~\cite{Yiyu2025,Yinji2025}.

Depending on the twist angle ($\theta$) between the two WSe$_2$ monoatomic layers, a moir\'e pattern emerges which leads to the creation of a mini-Brillouin zone and, most importantly, to flat electronic bands~\cite{Brzezinska_2025}. This points to a significant role played by electron-electron interactions. It is believed that tWSe$_2$ is in the moderately correlated regime, with the onsite Coulomb integral being comparable to the flat band width ($U\approx W$) \cite{Wang2020,Haining2020}. However, the actual strength of the electron-electron correlations is influenced by the twist angle and the dielectric environment of the sample. The experimental absence of the Mott insulating state at half-filling in Ref. \cite{Yinji2025} indicates that most probably a weak-to-moderate correlations regime ($U\lesssim W$) has been achieved in this particular study. Moreover, it has been established that the paired state emerges only in a relatively small range of displacement fields and band fillings, whose exact location depends on the twist angle.

The original experimental report as well as the most recent ones related with SC state in tWSe$_2$ have motivated theoretical analysis which considered effective single and multiband models as well as different forms of pairing mechanisms~\cite{Belanger2022_cDMFT_SC,Klebl2022,Chen2023,Zegrodnik2023, Wu_2023_PDW_SC,schrade2021nematic,Millis2023,Qimiao2025prl,Wei2024arxiv,Tuo2024arxiv, Millis2024_arxiv, DasSarma20225prb,Millis2025arxiv}. Many of the those considerations lead to a mixed singlet and triplet symmetry of the order parameter, due to the presence of the Ising type spin-orbit coupling in the system. Both the theoretical study and the experimental reports point to the role of the van Hove singularity in the stabilization of the paired state. However, the physical mechanism which determines the characteristic form of the phase diagram is still under ongoing debate and has not been completely resolved so far.

%in the $5.0^{\circ}$ twisted system

In our previous study, we have employed an effective single-band $t$-$J$-$U$ model to the description of SC state of tWSe$_2$, which lead to the stability of a topological mixed $d$+$id$ and $p$-$ip$ pairing scenario~\cite{Zegrodnik2023,Waseem2024}. Within such a description, the single particle physics is described by a tight binding Hamiltonian with complex hoppings which incorporate the spin-valley locking in the system. Here, we extend our previous considerations by supplementing the Hamiltonian with an {\em intersite Coulomb interaction term}. We focus on the effect of the displacement field in the formation of the paired state in order to discuss the obtained results in the view of the most recent experimental findings. In particular, our main aim here is to reconstruct one of the principal features of the tWSe$_2$, namely the appearance of an unconventional superconducting state in a relatively small range of displacement fields ($D$) and band fillings ($n$)~\cite{Yiyu2025,Yinji2025}. In our study, we apply the Gutzwiller approximation based on which we analyze in detail the electron correlation induced renormalization of particular energy terms in the context of the SC state stabilization. As we show explicitly, in the regime of weak-to-moderate correlations, the renormalization effects create favorable conditions for the SC pairing in close proximity of the half-filled situation. This effect, together with the van Hove singularity evolution induced by the displacement field determines the stability regime of the SC state in the ($n$,$D$) phase diagram.

\begin{figure}[!h]
 \centering
 \includegraphics[width=0.47\textwidth]{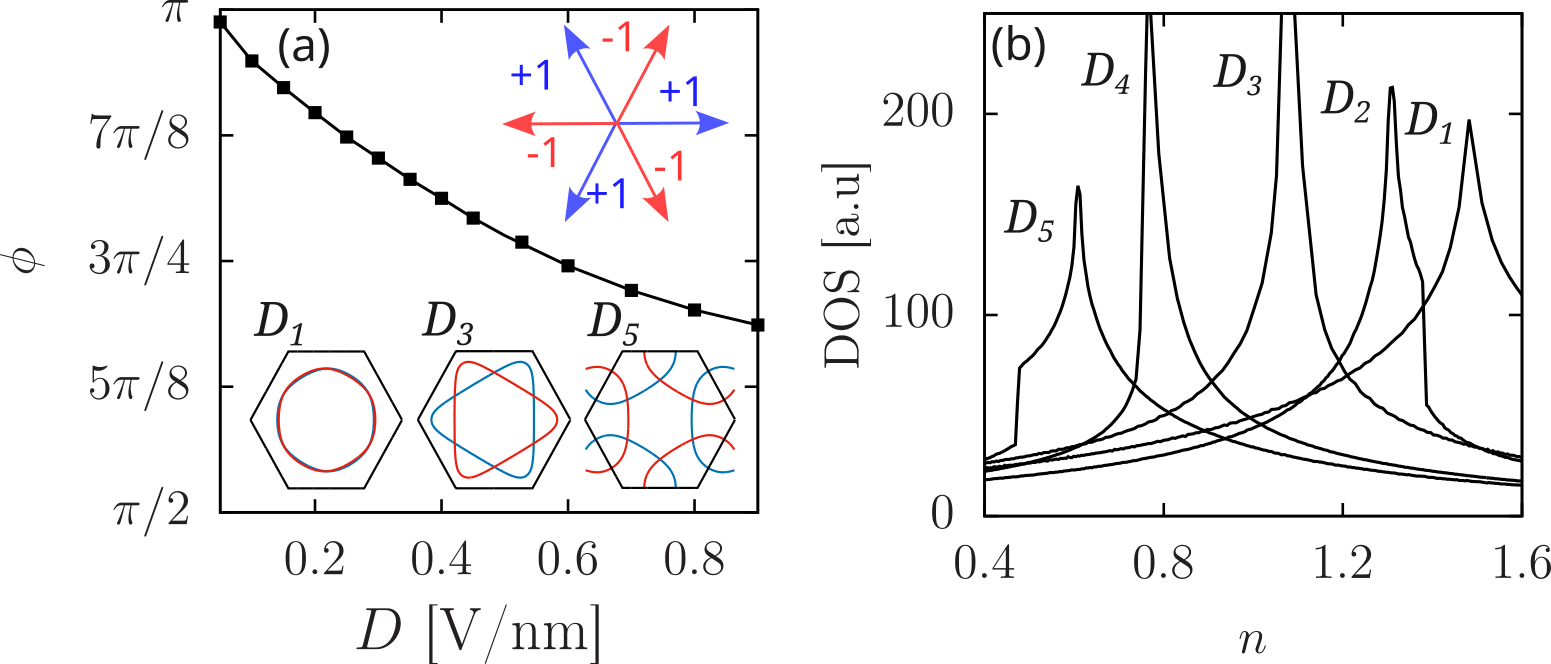}
 \caption{(a) Phase factor $\phi$ as a function of displacement field, $D$, together with spin-up (blue) and spin-down (red) Femi surfaces at half-filling for selected values of D (inset). In the upper right corner we provide the $\nu_{ij}=\pm 1$ factor corresponding to the six nearest neighbors [c.f. Eq. (\ref{eq:Hamiltonian_start})]. (b) Density of states at the Fermi level as a function of band filling for selected values of the displacement field $D_l\in\{0.05,~0.2,~0.3,~0.45,~0.6\}$~V/nm, for $l=1,2,3,4,5$, respectively.}
 \label{fig:phi_and_dos_Ddep}
\end{figure}

%%%%%%%%%%%%%%%%%%%%%%%%%%%%%%%%%%%%%%%%%%%%%%%%%%%%%%%%%%%%%%%%%%%%%%%%%%%%%%%%%%%%%%%
%%%%%%%%%%%%%%%%%%%%%%%%%%%%%%%%%%%%%%%%%%%%%%%%%%%%%%%%%%%%%%%%%%%%%%%%%%%%%%%%%%%%%%%
\textit{Model and method}.---We employ the moir\'e $t$-$J$-$U$ model supplemented with the intersite Coulomb repulsion term, i.e. 
\begin{equation}
    \begin{split}
    \hat{H}&=t\sum_{\langle ij\rangle\sigma}e^{i\sigma\nu_{ij}\phi}\;\hat{c}^{\dagger}_{i\sigma}\hat{c}_{j\sigma} \\
    &+ J\sideset{}{'}\sum_{\langle ij\rangle}\bigg(\hat{S}^z_i\hat{S}^z_j+\frac{1}{2}e^{i2\nu_{ij}\phi}\hat{S}^+_i\hat{S}^-_j+\frac{1}{2}e^{-i2\nu_{ij}\phi}\hat{S}^-_i\hat{S}^+_j\bigg)\\
     &+U\sum_{i} \hat{n}_{i\uparrow} \hat{n}_{i\downarrow} + V \sideset{}{'}\sum_{\langle ij\rangle} \hat{n}_{i} \hat{n}_{j},
    \end{split}
    \label{eq:Hamiltonian_start}
\end{equation}
where $\sigma=\pm 1$ represents spin up/down, $\hat{c}^{\dagger}_{i\sigma}$ and $\hat{c}_{i\sigma}$ are the creation and annihilation operators for electron with spin $\sigma$ at site $i$ of a triangular lattice, $\langle i,j\rangle$ correspond to nearest neighbors, $S^z_i$, $S^+_i$, $S^-_i$ are the spin-$\frac{1}{2}$ $z$ component, rising and lowering operators, respectively, $\hat{n}_{i\sigma}$ is the occupancy operator and $\hat{n}_{i}=\sum_{\sigma}\hat{n}_{i\sigma}$. The primed summation means that each bond between the lattice sites appears only once. The subsequent terms of the above Hamiltonian correspond to electron hopping, intersite exchange interaction, as well as intra-/inter-site Coulomb repulsion. The phase factors of the exponents have an alternating sign introduced by $\nu_{ij}=\pm 1$, which depends on the bond direction (cf. inset of Fig. \ref{fig:phi_and_dos_Ddep}). The resultant spin- and direction-dependent complex hoppings account for the Ising-type spin-orbit coupling in the system. As shown in Refs. \onlinecite{Wang2020,Haining2020}, the single particle part constitutes an effective description of the flat moir\'e band of twisted WSe$_2$/WSe$_2$ homobilayer. It should be noted that in the experimental situation the tWSe$_2$ structure is usually placed in between two electrods which allow to control both the electron concentration ($n$) and the displacement field ($D$) in an in-situ manner. The effect of the latter tunes the spin-valley splitting and enters the effective model via a $D$-dependent $t$ and $\phi$, which have been calculated for the twist angle of $\theta=5.08^{\circ}$ in Ref.~\onlinecite{Wang2020} and are used by us here. For the sake of completeness in Fig. \ref{fig:phi_and_dos_Ddep} we provide the $D$-dependence of $\phi$ as well as Fermi surfaces and density of states as a function of band filling for selected values of displacement field.

Motivated by the experimental result provided in Ref. \onlinecite{Yinji2025}, we focus on the weak-to-moderate correlations regime ($U/W\lesssim 1$) by taking $U=80$~meV, where the band width is $W\approx90$~meV. The value of the exchange interaction integral is set to $J=4t^2/U$. We apply the Gutzwiller approximation method which takes into account the electron-electron interactions above the level of a standard mean-field approach. At the same time, the considered method allows us to access the renormalization factors corresponding to subsequent energy terms appearing in our Hamiltonian and analyze them in the context of the SC state formation. The considered variational wave function of the Gutzwiller type has the following form $|\Psi_G\rangle=\hat{P}|\Psi_0\rangle$, where $|\Psi_0\rangle$ is the uncorrelated (mean-field) state, and $\hat{P}$ denotes the correlation operator, i.e., 
\begin{equation}
    \hat{P}=\prod_i\sum_{\Gamma}\lambda_{i,\Gamma}|\Gamma\rangle_{i\;i}\langle\Gamma|,
    \label{eq:correlation_operator}
\end{equation}
while $|\Gamma\rangle$ are states from the local basis $\{ |\emptyset\rangle,\;|\uparrow\rangle,\;|\downarrow\rangle,\;|\uparrow\downarrow\rangle \}$, and $\lambda_{i,\Gamma}$ represent variational parameters determined via energy minimization. Since we consider a homogeneous system without magnetic or charge ordering, we take $\lambda_{i,\uparrow\downarrow}\equiv \lambda_{\uparrow\downarrow}$, $\lambda_{i,\uparrow}=\lambda_{i,\downarrow}\equiv\lambda_{s}$, $\lambda_{i,\emptyset}\equiv \lambda_{\emptyset}$.

%$\lambda_{i,\Gamma}\equiv \lambda_{\Gamma}$ and $\lambda_{\uparrow}=\lambda_{\downarrow}\equiv \lambda_{s}$. 
\begin{figure*}[!t]
\centering
\includegraphics[width=1.0\textwidth]{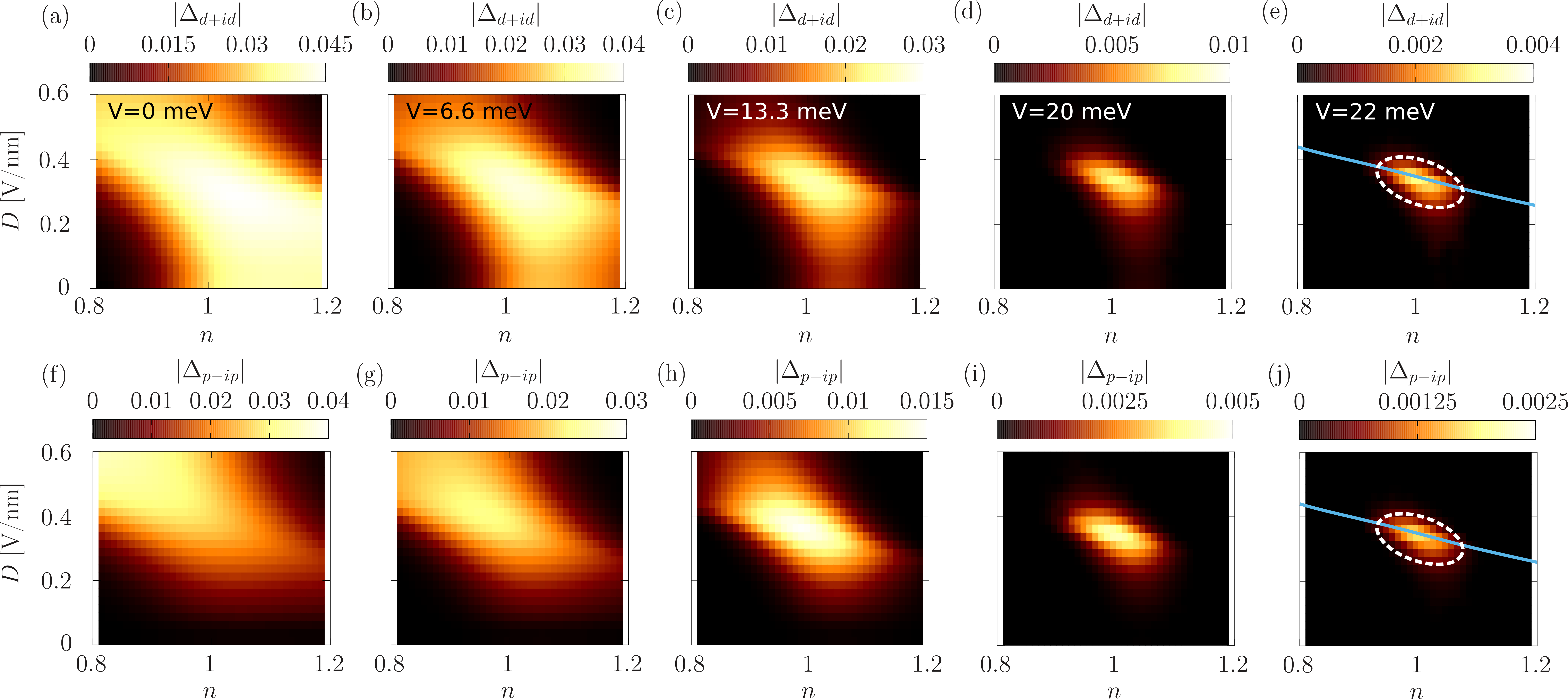}
\caption{Symmetry resolved superconducting gap amplitudes corresponding to the obtained mixed spin-singlet $d+id$ and spin-triplet $p-ip$ paired state as a function of band filling $n$ and and displacement field $D$ for gradually increasing value of the intersite Coulomb repulsion $V$. The onsite Coulomb repulsion and the exchange interaction parameters are set to $U=80$\;meV and $J=4t^2/U$, respectively. The solid blue line in (e) and (j) mark the evoulution of the van Hove singularity across the phase diagram. The white dashed line is a guide to the eye denoting the stability range of the SC state.}
\label{fig:diagrams_DnV_dep}
\end{figure*}

It is important to note that, when it comes to stabilization of the superconducting state, the exchange interaction term ($J$) and the intersite Coulomb interaction term ($V$) have positive and negative effects, respectively. Within the Gutzwiller approximation the expectation values in the $|\Psi_G\rangle$ state of the two mentioned terms is the following
\begin{equation}
\begin{split}
    E_J&=\lambda^4_sJ\;\sideset{}{'}\sum_{\langle ij\rangle}\bigg(\frac{1}{2}\;\sum_{\sigma}\;e^{i\sigma2\phi_{ij}}\langle\hat{c}^{\dagger}_{i\sigma}\hat{c}_{i\bar{\sigma}}\hat{c}^{\dagger}_{j\bar{\sigma}}\hat{c}_{j\sigma} \rangle_0\\
    &+\frac{1}{4}\;\sum_{\sigma\sigma'}\sigma\sigma'\big(\langle\hat{n}_{i\sigma}\hat{n}_{j\sigma'}\rangle_0-n_s^2\big)\bigg),\\
    E_V&=V \sideset{}{'}\sum_{\langle ij\rangle}\big( g^2_v\langle \hat{n}_i\hat{n}_j\rangle_0-(1-g^2_v) n_s^2\big),
\end{split}
\label{eq:J_V_expectation_value}
\end{equation}
where $g_v=(1-(2-\lambda_{\uparrow\downarrow})n_s)/(1-n_s)$, $n_{i\uparrow}=n_{i\downarrow}\equiv n_s$, $n_{i\sigma}=\langle \hat{n}_{i\sigma}\rangle_0$, and $\langle\hat{o} \rangle_0$ referring to an expectation value of operator $\hat{o}$ in the $|\Psi_0\rangle$ state. As one can see, the expectation value of a given energy term in the correlated state, $|\psi_G\rangle$, can be expressed in terms of the expectation values in the uncorrelated state, $|\psi_0\rangle$. However, factors $\lambda_s^4$ and $g_v^2$ renormalize the coupling constants $J$ and $V$, respectively. Therefore, the relative balance between the two energy terms is going to be affected by the significant onsite Coulomb repulsion via the renormalization factors possibly having a decisive influence on the stabilization of the superconducting state.

To determine the values of the real-space superconducting gap amplitudes in the correlated state, $\Delta_{ij}^{\sigma\bar{\sigma}}=\langle\hat{c}^{\dagger}_{i\sigma}\hat{c}^{\dagger}_{j\bar{\sigma}}\rangle_G$ , we apply the Effective Hamiltonian scheme ~\cite{Kaczmarczyk2015} and then extract the symmetry resolved SC gaps~\cite{Zegrodnik2023} which allow us to analyze the principal features of the paired state. Also, following B\"unemann et al. we impose an additional condition to the correlation operator, which reduces the complexity of the numerical calculations~\cite{Bunemann_2012}. More details related to the applied theoretical approach are provided in the SM\;\cite{Zegrodnik_SM}. In the following, we refer to the average number of electrons per lattice site as band filling $n=2n_s$.

%%%%%%%%%%%%%%%%%%%%%%%%%%%%%%%%%%%%%%%%%%%%%%%%%%%%%%%%%%%%%%%%%%%%%%%%
\textit{Results}.---In order to discuss our theoretical approach in the context of the available experimental data, we calculate the symmetry-resolved SC amplitudes across the $(n,D)$-diagram. In Fig. \ref{fig:diagrams_DnV_dep} we provide the results obtained for the gradually increasing value of the intersite Coulomb repulsion, $V$. Our calculations lead to the stabilization of a mixed $d$+$id$ (singlet) and $p$-$ip$ (triplet) paired state. Singlet-triplet mixing has been also reported in other theoretical descriptions of tWSe$_2$\;\cite{Klebl2022, Millis2025arxiv,Sheng2023,Millis2024_arxiv,schrade2021nematic} and is a straightforward consequence of the spin-valley locking as we discuss in more detail in the SM\;\cite{Zegrodnik_SM}. As one can see, for the $V=0$ case the SC state is relatively robust, with the spin-singlet pairing dominating the low displacement field regime and gradually increasing contribution resulting from the triplet component as $D$ increases. Such robust SC state covering a significant area of the phase diagram is in contradiction with the experimental situation. However, as the value of $V$ is increased, the stability area of the paired state shrinks. It should be noted that the negative effect of the $V$-term on the pairing has been previously reported in other theoretical considerations\;\cite{Kivelson_2012,Thomale_2012,Thomale_Chubukov_2014}. For $V=22$~meV the SC phase covers only a small part of the $(n,D)$ plane which is located in the proximity of the half-filled situation, and for $D\approx0.2-0.4$~V/nm. The SC phase shown in Fig. \ref{fig:diagrams_DnV_dep} (e) and (j) is characterized by a Chern number $C=\pm 4$ indicating a topological character. The sign change of $C$ appears exactly at the Van Hove singularity line which is due to the transition from the electron-like to hole-like Fermi surfaces. For more details related with the topological features, see the SM\;\cite{Zegrodnik_SM}. It should be noted that a similar form of the phase diagram has recently been reported experimentally for tWSe$_2$ in Ref. \cite{Yinji2025}. However, due to  uncertainty in experimental determination of band filling, it is not completely clear if the SC phase stability range has reached the half-filled situation. Another experimental report has shown the SC phase stability at half-filling but in very low displacement field range~\cite{Yiyu2025}.

\begin{figure}[!h]
 \centering
 \includegraphics[width=0.47\textwidth]{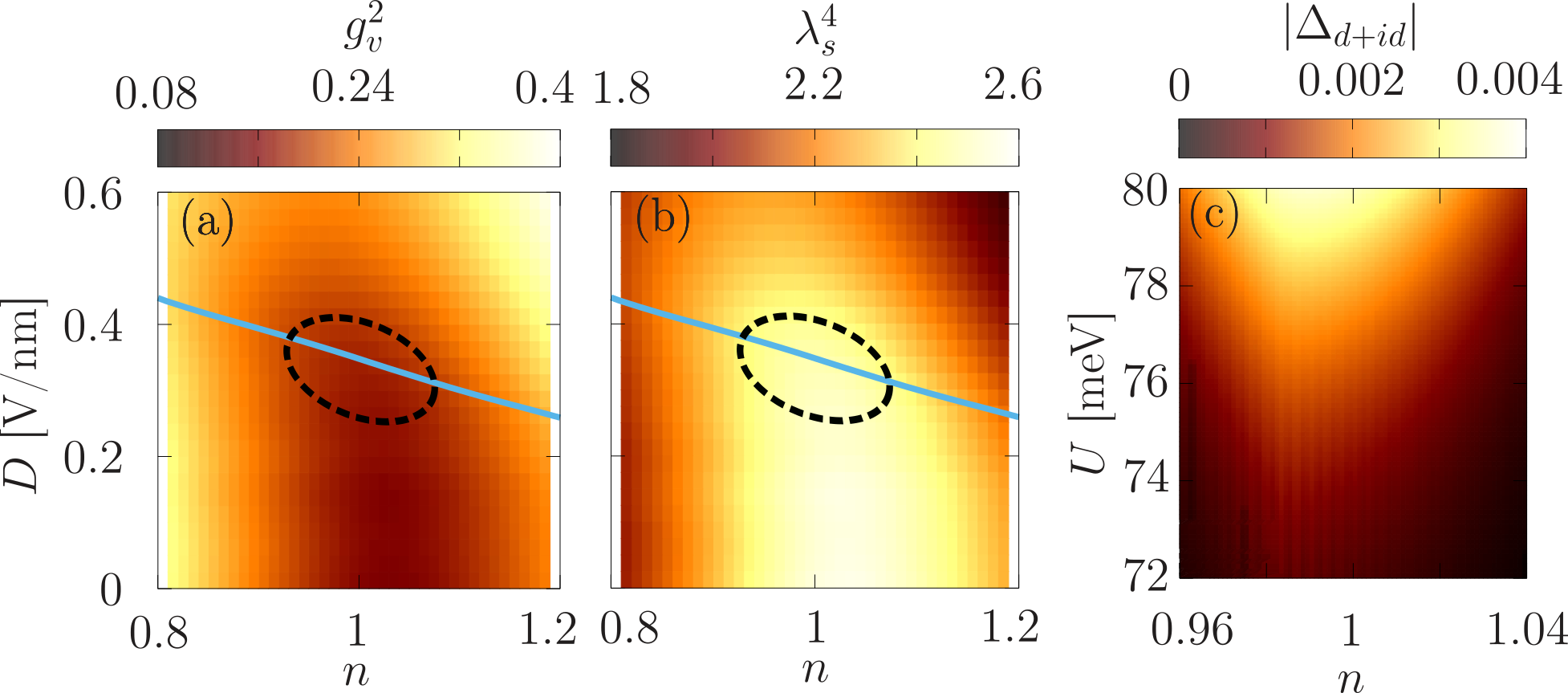}
 \caption{The $g_v^2$ (a) and $\lambda_s^4$ (b) factors which renormalize the intersite Coulomb interaction term and the exchange interaction term [cf. Eq. (\ref{eq:J_V_expectation_value})], respectively, as a function of $n$ and $D$. The blue line marks the Van Hove singularity evolution across the phase diagram. The black dashed line correspond to the stability of the SC state as in Fig. \ref{fig:diagrams_DnV_dep} (e,j). Note, that SC phase is located at the crossing of the Van Hove singularity line and the area where the renormalization is the strongest. (c) The SC gap as a function of $U$ and $n$ for $D=0.35$~V/nm with $J=4t^2/U$, $V=U/3.635$. We provide only the singlet component since the triplet one shows the same behavior but is approximately twice smaller.}
 \label{fig:renorm_params}
\end{figure}

In order to analyze why the particular small area of the $(n,D)$-plane seen in Fig. \ref{fig:diagrams_DnV_dep} (e) and (j), creates favorable conditions for the SC state to be stable, we first study the evolution of Van Hove singularity across the phase diagram. The solid blue line marks the band fillings which for a given value of $D$ correspond to the Van Hove singularity being located at the Fermi level. As one can see, the area covered by the SC state is located around the Van Hove singularity line. Such an effect is expected since high values of the density of states at the Fermi energy stabilize superconductivity in the weak-coupling theory. The second effect which is of significant importance here originates from the electron-electron correlations. In Figs. \ref{fig:renorm_params} we provide the ($n$,$D$)-dependence of the $g_v^2$ and $\lambda_s^4$ factors, which renormalize the intersite Coulomb interaction term and the exchange interaction term, respectively [cf. Eq. (\ref{eq:J_V_expectation_value})]. For $g_v^2=\lambda_s^4\equiv 1$ our variational Gutzwiller state would become equivalent to the Hartree-Fock solution ($|\psi_0\rangle=|\psi_G\rangle$), which is not the case here. More importantly, the region of maximized renormalization is located roughly around half-filling where $g_v^2$ is small, significantly diminishing the destructive character of the $V$ term on the paired phase. Moreover, $\lambda_s^4$ is relatively large in the same region, which enhances the positive effect of the $J$-term on the pairing. At the same time, since we are dealing with the weak-to-moderate correlation regime, there is no Mott insulating state located at half-filling which would suppress the SC state there. As a result, the stability of the SC state lies exactly at the crossing of the Van Hove singularity line and the maximized renormalization area at half-filling. This allows for the two positive effects to act simultaneously which as a result create favorable conditions for the paired phase only in a particular region of the ($n$,$D$)-plane. Additionally, in Fig. \ref{fig:renorm_params} (c) we show that the SC phase vanishes while the electron-electron correlations are weakened.

\begin{figure}[!h]
 \centering
 \includegraphics[width=0.47\textwidth]{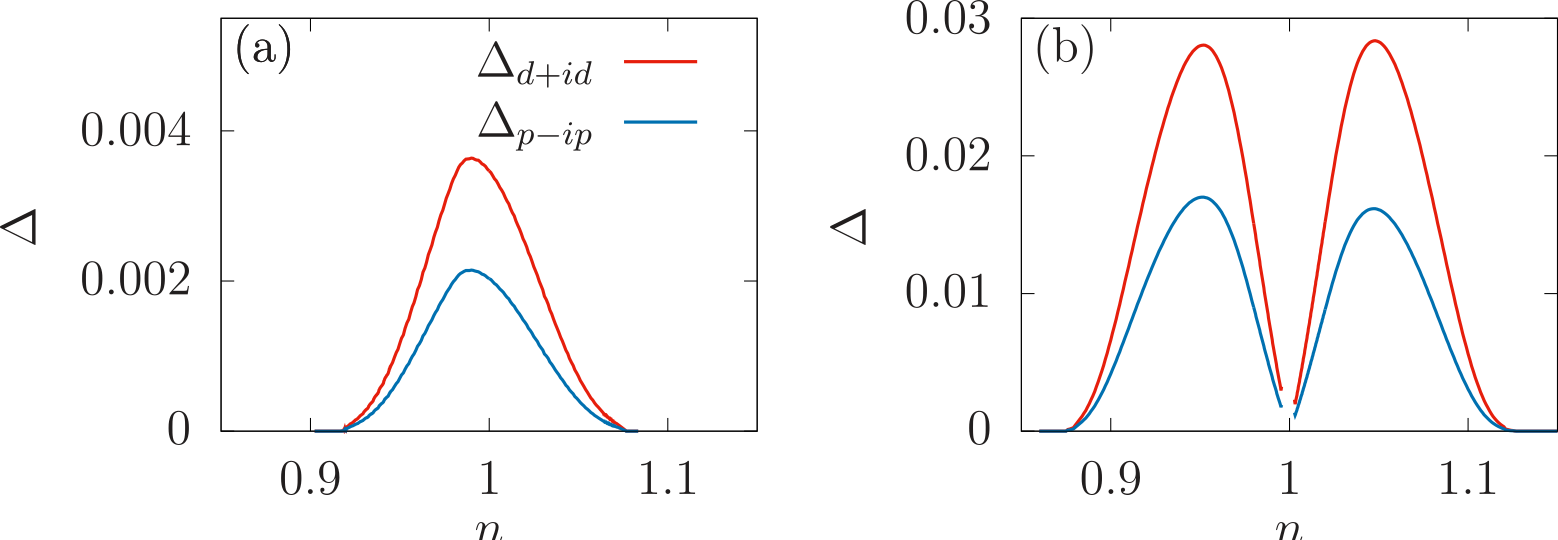}
 \caption{Symmetry resolved gap amplitudes of the mixed singlet-triplet solution as a function of $n$ for $D=0.35$~V/nm and for two selected values of the onsite Coulomb repulsion, $U=80$~meV (a) and $U=120$~meV (b). In both cases we keep $J=4t^2/U$ and $V=U/3.635$.}
 \label{fig:ndep}
\end{figure}

For comparison, in Fig. \ref{fig:ndep} we provide the results for both weak-to-moderate and strong correlations. Namely, (a) and (b) correspond to $U=80$~meV and $U=120$~meV, respectively, while the band width is $W\approx 90$~meV. In the second situation, instead of having a single SC dome, we actually obtain two of them with the SC gap reaching values one order of magnitude larger than in the first case. This is due to the fact that for $U\gtrsim W$ strong electron-electron correlations enhance the positive effect of the renormalization parameters $g_v^2$ and $\lambda^4_s$ on the pairing. However, due to the fact that $U$ is already larger than $W$ in (b), a Mott insulating state starts to develop at half-filling which suppresses the SC amplitudes there, leading to a two-dome behavior. 

It should be noted that recent measurements carried out for $\theta=5.0^{\circ}$ and $\theta\approx3.5^{\circ}$ twisted WSe$_2$ show a single SC dome not separated by a Mott state in a certain $D$-range~\cite{Yinji2025,Yiyu2025}. However, the original experimental report referring to the same system with a twist angle $\theta=5.1^{\circ}$ shows signatures of two SC domes and a Mott insulator at half filling~\cite{Wang2020}. It might be possible that depending on the dielectric environment and the actual twist angle the two regimes ($U\lesssim W$ and $U\gtrsim W$) can be reached in tWSe$_2$ leading to two different forms of the phase diagram. It should be noted that our results lead to a significantly larger maximal SC gap in the two-dome case, which would agree with the experimental situation where $T_C\approx200$-$300$~mK and $T_C\approx1$~K correspond to single-dome and two-dome behaviors, respectively~\cite{Wang2020,Yinji2025,Yiyu2025}. 

During the resubmission process of this article a new experimental report has appeared\;\cite{xia2025_arxiv} showing the stability of the SC phase close to the crossing between the van Hove singularity and the half-filling, which is consistent with our approach. However, above a certain value of $D$, an antiferromagnetic Mott insulator begins to develop at $n=1$.

Our Gutzwiller approach is limited to the ground state only. Therefore, we are not able to reach non-zero temperatures. Nevertheless, in order to provide a rough estimate of the maximal critical temperature in the two considered situations, we use the formulas: $\Delta(T=0)=1.764k_BT_C$ and $\Delta(T=0)=J\Delta_{\mathrm{max}}$, with $\Delta_{max}$ being the maximal SC gap amplitude value obtained in our calculations. Such an estimate gives us $T_C\approx 100$~mK and  $T_C\approx 450$~mK for the $U=80$~meV (a) and $U=120$~meV (b), respectively. The obtained values underestimate the experimentally obtained critical temperatures; however, they are of the same order.

%%%%%%%%%%%%%%%%%%%%%%%%%%%%%%%%%%%%%%%%%%%%%%%%%%%%%%%%%%%%%%%%%%%%%%%%5

\textit{Final remarks}.---
We have analyzed the principal features of the unconventional superconducting state in tWSe$_2$ within an effective $t$-$J$-$U$ model supplemented with the intersite Coulomb repulsion term. As we show, for the weak-to-moderate correlations regime the topological mixed singlet-triplet superconducting state covers a relatively small area of the ($n$,$D$) phase diagram which is located where the Van Hove singularity line crosses half-filling. 
We show that this precise region optimizes the conditions for superconductivity: the VHS causes a large density of states, whereas the electron-correlation-induced renormalization is maximized at half-filling. This allows the two effects to act simultaneously, creating favorable conditions for the paired state as seen in experiments\;\cite{Yinji2025,Yiyu2025,xia2025_arxiv}. To explore the unconventional nature of the SC state experimentally, we propose using Knight shift measurements to identify any non-singlet pairing. Observing the Kerr effect could indicate spontaneous breaking of time-reversal symmetry. Lastly, our prediction of topological superconductivity in tWSe$_2$ highlights it as a promising platform for detecting chiral edge states through scanning techniques or transport measurements.\cite{Qi.2011}

Additionally, we show that a single SC dome and two SC domes can be reproduced within a single theoretical framework. The former corresponds to weaker correlations and the latter to stronger correlations. The two-dome scenario comes as a result of the fact that the Mott insulating state located at half-filling separates the SC stability range into two. Nevertheless, the maximal values of the gap amplitudes are larger for the stronger correlations and the two-dome scenario. This is in qualitative agreement with the available experimental data~\cite{Wang2020,Yinji2025,Yiyu2025}.

It would be interesting to analyze the effect of displacement field on SC in tWSe$_2$ by using a description based on the Hubbard model in weak-to-moderate correlations. For that it would be necessary to apply a method which takes into account the correlations effects with better accuracy like e.g. DMRG. We should see progress along this line soon~\cite{Biborski2025NEW}.

\textit{Acknowledgement}.---This research was funded by National Science Centre, Poland (NCN) according to decision 2021/42/E/ST3/00128. We gratefully acknowledge Polish high-performance computing infrastructure PLGrid (HPC Center: ACK Cyfronet AGH) for providing computer facilities and support within computational grant no. PLG/2024/017372. LR is funded by the Swiss National Science Foundation through Starting Grant No. TMSGI2\_211296. The code that was written to perform the numerical calculations and the data behind the figures are available in the open repository~\cite{zegrodnik_michal_2025_zenodo}. For the purpose of Open Access, the author has applied a CC-BY public copyright license to any Author Accepted Manuscript (AAM) version arising from this submission.

\bibliography{refs.bib}

\end{document}